# Evaluation of head segmentation quality for treatment planning of tumor treating fields in brain tumors

Reuben R Shamir[1] and Zeev Bomzon[1]

[1] Novocure, Haifa, Isreal
`rshamir@novocure.com`

## 1    Purpose

Tumor treating fields (TTFields) is an FDA approved therapy for the management of Gliobastoma Multiform (GBM) and under study for multiple additional indications [1]. It involves the continuous placement of transducer arrays (TAs) on the patient's head and delivery of an electric field to the tumor area. Larger electric field intensities within the tumor are associated with improved treatment outcomes. The electric fields intensities directly depend on the location of the TAs. We have developed a semi-automatic protocol for estimating the electric field within the tumor of a specific GBM patient and TAs layout [2]. The method is composed of three key steps: 1) head segmentation and the assignment of electrical tissue properties (conductivity and permittivity); 2) virtual placement of the TAs on the outer surface of the head, and; 3) simulation of the electric field propagation and estimation of the dose within the tumor. The head segmentation was performed semi-automatically, first using SPM-MARS [3] and tweaking its parameters, and then by manually fixing the segmentation errors. As we move towards large scale planning software, we have developed an in-house atlas-based automatic head segmentation method. To ensure that estimates of TTFields dosage remain similar and relevant to the outcomes, we developed a process for estimating the quality of the new segmentation method. Our method was specifically designed for evaluating atlas-based segmentation algorithms, aiming to facilitate a better estimation. More general methods [4–6] can also be incorporated for this purpose.

## 2    Methods

We present a method for predicting a similarity measure between computed- and validated- segmentations of the head, but in the absence of the validated one (Fig. 1). To measure the quality of the new segmentation method, the Dice coefficient was measured between the computed- and validated- head segmentations of our training set (see below). Then, we investigated four categories of features that seem to be relevant for atlas based segmentation methods: 1) quality of global (affine) registration; 2) quality of local (deformable) registration; 3) input image properties, and; 4) geometrical properties of the segmented tissues. Specifically, *global registration quality* is estimated with inverse consistency [7]. We have extracted the following features to account for the *deformable registration quality*: 1) deformation's field bias (mean of all vectors); 2) directional variability (SD of the 3-element vector that is the mean of each axis), and; 3) mean per-axis variability (mean of 3-element vector that is the SD of each axis). We have incorporated the shortest axis length and the signal to noise ratio of each tissue separately as *input image quality*. Many features that describe the segmentation shape can be defined. In this study, we have selected two simple measures: the volume of the shape and the number of connected components. These measures were computed per tissue (Fig. 2a).

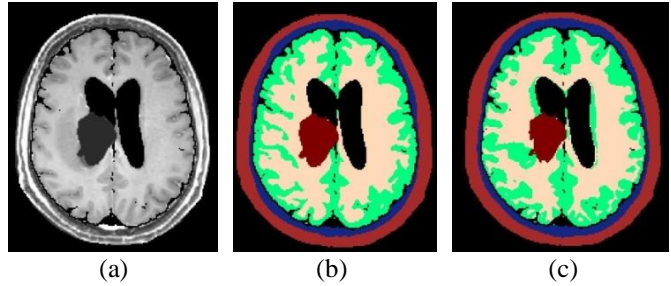

(a)  (b)  (c)

**Fig. 1.** (a) MRI T1w of a GBM patient's head; (b) the segmentation of a recently developed algorithm, and; (c) a validated segmentation. The tumor was semi-automatically pre-segmented in both segmentations. The goal of this study is to predict the similarity between (b) and (c) given (a) and (b), but in the absence of (c).

The features were incorporated in a decision tree regressor [8]. We applied a leave-one-out approach on 20 TTFields patients' head MR-T1 images, their validated segmentations and their automatically generated counterparts. We compared the measured Dice coefficients between the sets to the Dice coefficients' predictions.

## 3  Results

Fig. 2a presents the observed absolute correlation values. The Dice values of cerebrospinal fluid (CSF) segmentations were significantly ($p < 0.05$) correlated with registration consistency, shortest axis length and skin's number of connected components (NCC). The skin and muscle were both labeled as 'skin' since they share similar electrical properties. Their Dice coefficients were significantly correlated with the signal to noise ratio (SNR) in these tissues on the MRI image. The grey matter (GM) Dice values were significantly correlated with registration consistency, deformation (bias, direction and variability), shortest axis length, volume of white matter (WM) and NCC of CSF, skin, and GM. The WM Dice values were significantly correlated with the registration consistency and deformation bias, shortest axis length, the volume of CSF and WM, and NCC of CSF, Skin and GM. The skull Dice values were significantly correlated with the SNR of this tissue on the MRI image. The predictions of the decision tree regressor were similar and highly correlated with the actual ones (average absolute difference 3% (SD = 3%); r = 0.92, p < 0.001; Fig. 2b).

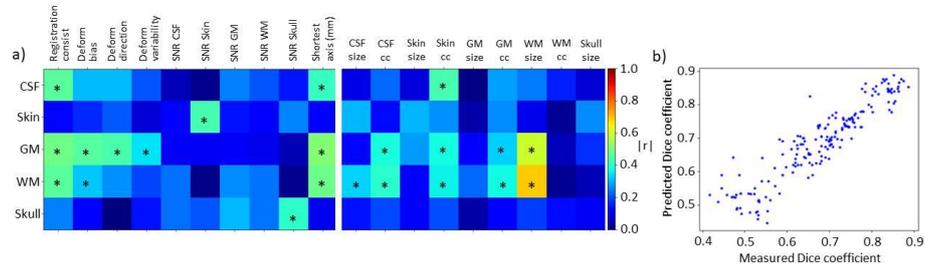

Fig. 2. (a) Absolute Pearson's correlation between computed features (columns) and the Dice coefficients of each segmented tissue (rows). The Dice coefficients were computed between the validated head segmentations and those that were computed with a new automatic segmentation method (* p < 0.05). (b) The suggested features and a decision tree regressor output predictions of Dice coefficients that are in a high correlation with the actual ones (r = 0.92; p < 0.001). CSF - cerebrospinal fluid; Skin - skin and muscle; GM - grey matter; WM - white matter; SNR – signal to noise ratio; cc - # connected components.

## 4    Conclusions

Our results suggest that, for the specific algorithm that was tested, intra-cranial tissues segmentation's accuracy is related to registration quality (global and local) and to the shortest axis length (Fig. 2a). In contrast, extra-cranial structures are related to image quality (SNR). Moreover, the above features can be incorporated in a regression method to predict the expected Dice coefficient. Therefore, we conclude that segmentation's quality estimation is feasible by incorporating a machine learning approach and features that are relevant to the segmentation.

While this study was limited to a specific segmentation method, a similar methodology can be applied to other methods and facilitate an in depth understanding of their pros and cons. For example, our results suggest that, as expected, the atlas-based segmentation method is highly depending on accurate atlas-to-patient registration.

Another limit of this study is that Dice coefficient is only one parameter related to accuracy, but does not necessary represent the actual quality of the segmentation. Moreover, it was suggested that the Dice is related to tissue's volume. Therefore, we plan to study other measures and examine the effect of tissue's volume on our results.

Currently we are investigating the effects of segmentation errors on the TTFields simulation results to characterize the required accuracy in our application. Moreover, we consider incorporating the simultaneous truth and performance level estimation (STAPLE) method [5] or one of its variants [6, 9] to improve the prediction of segmentation quality.